%% file: main.tex
\documentclass{interspeech2024}

\interspeechcameraready
\usepackage{multirow}
\usepackage[table,xcdraw]{xcolor}
\usepackage{paralist}
\usepackage[font=small,skip=0pt]{caption}
\usepackage{mathptmx}
\usepackage{tikz}

\title{DAISY: Data Adaptive Self-Supervised Early Exit \\ for Speech Representation Models}
\name[affiliation={1}]{Tzu-Quan}{Lin}
\name[affiliation={1}]{Hung-yi}{Lee}
\name[affiliation={2}]{Hao}{Tang}
\address{
  $^1$Graduate Institute of Communication Engineering, National Taiwan University, Taiwan\\
  $^2$University of Edinburgh, United Kingdom}
\email{tzuquanlin@gmail.com}
\begin{document}
\maketitle 
\begin{abstract}
Self-supervised speech models have shown to be useful for various tasks, but their large size limits the use in devices with low computing power and memory. In this work, we explore early exit, an approach for reducing latency by exiting the forward process of a network early. Most approaches of early exit need a separate early exit model for each task, with some even requiring fine-tuning of the entire pretrained model. We introduce \textbf{D}ata \textbf{A}dapt\textbf{i}ve \textbf{S}elf-Supervised Earl\textbf{y} Exit (\textbf{DAISY}), an approach that decides when to exit based on the self-supervised loss, eliminating the need for multiple round of training and fine-tuning. DAISY matches the performance of HuBERT on the MiniSUPERB benchmark, but with much faster inference times. Our analysis on the adaptivity of DAISY shows that the model exits early (using fewer layers) on clean data while exits late (using more layers) on noisy data, dynamically adjusting the computational cost of inference based on the noise level of each sample.
\end{abstract}

\input{introduction}
\input{related}
\input{methodology}
\input{experiments}
\input{results}
\input{analysis}
\input{conclusion}
\input{adknowledgement}

\bibliographystyle{IEEEtran}
\bibliography{ref}

\end{document}

%% file: introduction.tex
\section{Introduction}

Self-supervised speech models \cite{chung2019unsupervised, liu2020mockingjay, baevski2020wav2vec, hsu2021hubert, chen2022wavlm, chung2021w2v} have demonstrated impressive capabilities in feature extraction. The feature extracted by speech SSL models could generalize well across various downstream datasets \cite{shu2021superb, tsai2022superb, 2022superbslt}. However, to effectively learn contextualized information from unsupervised data, these models are typically large \cite{zhang2022bigssl}, leading to substantial computational costs during inference. This significantly restricts the use of speech SSL models on low-resource devices.

To solve this issue, many studies have focused on compressing self-supervised speech models. Prior work has explored knowledge distillation \cite{chang2022distilhubert, wang2022lighthubert,
lee2022fithubert} to train a smaller student model to mimic the behavior of a larger teacher model. Some use either unstructured \cite{lai2021parp} or structured \cite{peng2023structured, wang2023task} pruning to remove redundant parameters in self-supervised models, and some apply quantization methods \cite{peng2021shrinking} to store the models at lower bitwidths. Early exit \cite{teerapittayanon2016branchynet}, unlike the other methods mentioned, does not reduce the number of parameters in the model, but instead aims to stop the forward process to make predictions in earlier layers, thereby reducing inference latency.

In this work, we will focus on the early exit approach.
Most studies on early exit for self-supervised models are for processing texts \cite{xin2020deebert, zhou2020bert, xin2021berxit, li2021accelerating, schuster2022confident, hu2023smartbert}, with only a few applied to speech \cite{yoon2022hubert}. Typically, these methods involve attaching early exit branches to each layer and fine-tuning these branches on specific downstream dataset. During inference, the decision to exit at a particular layer is made by assessing the entropy or confidence of the branch \cite{teerapittayanon2016branchynet}. Existing approaches require training early exit branches for every downstream tasks, and many require fine-tuning the self-supervised models for specific downstream tasks as well, which significantly increases the computational cost. Moreover, the effectiveness of early exit has not been shown other than ASR \cite{yoon2022hubert}.

In this work, inspired by the finding that the self-supervised loss aligns well with downstream performance \cite{lin2022compressing}, we propose \textbf{D}ata \textbf{A}dapt\textbf{i}ve \textbf{S}elf-Supervised Earl\textbf{y} Exit (\textbf{DAISY}), a novel and effective method using the entropy of the self-supervised loss to make decisions for early exit. Our approach can be divided into three stages. In the first stage, we attach an early exit branch to each hidden layer and optimize each branch with a self-supervised objective (e.g., predicting quantized representation as in HuBERT \cite{hsu2021hubert}). In the second stage, we fix the early exit branches, and train classifiers for downstream tasks using the weighted sum of features following SUPERB \cite{shu2021superb}. Different from typical early exit methods that only perform early exit at the inference phase, we incorporate early exits during training of the downstream classifier. In the third stage, all models are frozen for inference with early exit. Note that the self-supervised model is frozen during all three stages. Compared to previous early exit methods, our approach requires training early exit branches only once in the first stage, and the early exit branches are then used for \emph{all} downstream tasks. Our approach does not require fine-tuning the self-supervised model either, significantly more efficient in training than other approaches. 

Since our approach no longer depends on the downstream tasks, when and where a branch decides to exit solely depends on properties of the input data.
In other words, regardless of the task, once the input data is given, the model will exit at the same layer.
What decides when to exit is not the task, but the characteristics of the data set on which the task is performed.
This prompts us to study the data adaptivity of our approach.
Based on our analysis with different noise levels, models tend to forward to deeper layers on data with low signal-to-noise ratio. This allows it to achieve a favorable performance and speed-up trade-off on datasets with varying degrees of noise.

%% file: related.tex
\section{Related Work}

The concept of early exit was first proposed by BranchyNet \cite{teerapittayanon2016branchynet}, where they proposed adding early exit branches to the intermediate layers of the model. During inference, it is possible to exit early via these branches when labels can be inferred with high confidence. 

DeeBERT \cite{xin2020deebert} introduced the concept of early exit to self-supervised models. Similar to BranchyNet, DeeBERT attaches branches to intermediate layers of BERT and fine-tunes these branches for specific downstream tasks. During inference, it determines whether to perform an early exit by assessing the confidence of the predictions.
CALM \cite{schuster2022confident} proposed to use token-level early exit to accelerate the process of language model generation. They validate the feasibility of early exit in generation tasks. However, they still need to train the early exit branches for different downstream tasks.

HuBERT-EE \cite{yoon2022hubert} utilizes early exit to accelerate the inference process of HuBERT. However, they only validate their approach on ASR, and similar to all previous early exit methods, their method requires training branches for different downstream tasks. In addition, they need to fine-tune the entire HuBERT model first before training early exit branches.

%% file: methodology.tex
\section{Methodology}

\begin{figure*}
  \begin{center}
  \includegraphics[width=\linewidth]{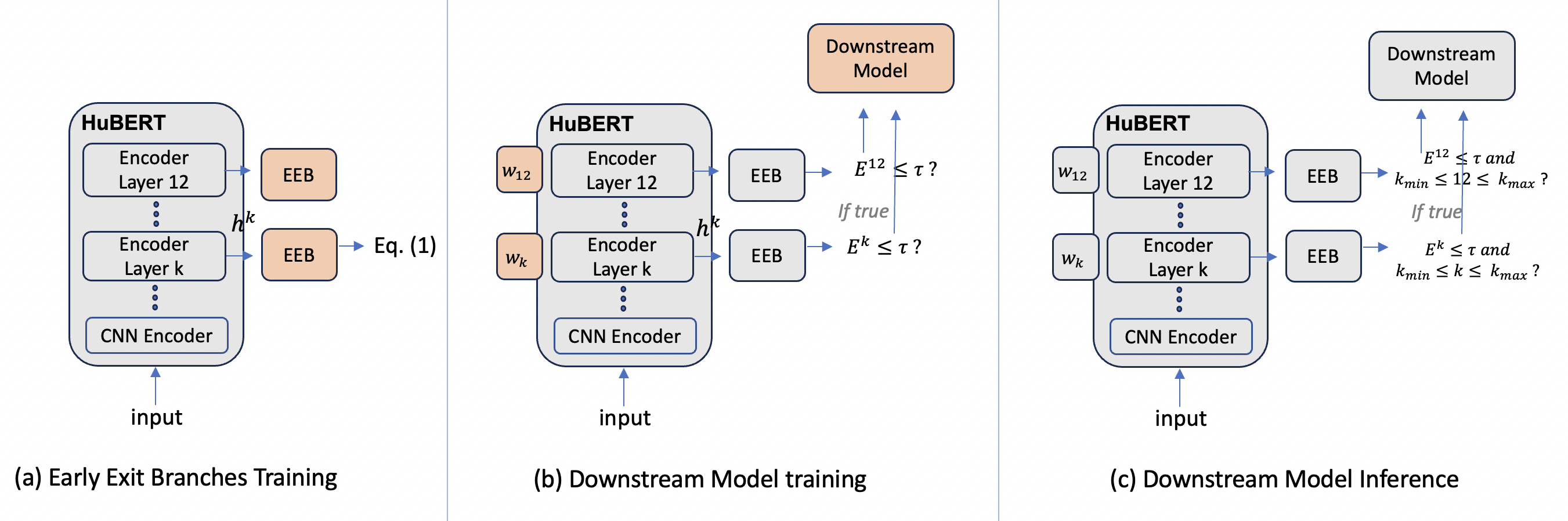}
  \end{center}
  \caption{Three stages of DAISY: (a) the training of early exit branches, (b) the training of downstream models, and (c) early exit at inference time. EEB is used to denote early exit branches, where the linear classifier and entropy computation happen. Model parameters are frozen when the boxes are in gray; model parameters are being trained when the boxes are in orange.
   \label{fig:DAISY}}
\end{figure*}

In essence, early exit amounts to learning a function (a branch) $f^k$ for layer $k$ such that, when taking the $k$-th hidden layer $h^k$ as input, if $f^k(h^k) = 1$, the forward process stops and the model exits early; otherwise, if $f^k(h^k) = 0$, the forward process continues.
In this section, we will introduce our approach, DAISY, and discuss how each early exit branch is trained. There are three distinct phases of DAISY, and they are summarized in Figure \ref{fig:DAISY}.

\subsection{Training early exit branches}

Instead of training early exit branches with downstream tasks, we propose to train them with a self-supervised loss.
We choose a simplified variant of the HuBERT loss, measuring the cross entropy between a quantized frame and the output of a linear classifier.
Formally, for every layer $k$, we train a linear classifier $W^k$ with
\begin{align}
    \frac{1}{T} \sum_{t=1}^{T} CE(W^k h^k_t, y_t),
\end{align}
where $h^k_t$ is the hidden vector at layer $k$ and time $t$, and $y_t$ is the target at time $t$.
The target $y_t$ should ideally be the targets used for training HuBERT,
but the targets are never released.
We instead create targets by running HuBERT and taking the argmax from the last linear layer.
This variant of HuBERT loss have been shown to be as effective as the original HuBERT loss \cite{lin2022compressing} and is computationally more efficient \cite{yang2023fast}.

After training the early exit branches, given a data sample, we compute the entropy of the $k$-th branch 
\begin{align}
    E^k = \frac{1}{T}\sum_{t=1}^{T}\sum_{c=1}^{C}-p^k_{t,c}ln(p^k_{t,c}),
\end{align}
where $p^k_{tc}$ is the probability of the class $c$ at layer $k$ and time $t$ obtained with the linear layer $W^k$ and a softmax.
The entropy is by no means the perfect measure of uncertainty or confidence (see \cite{ovadia2019can} and the citations therein), but it is simple to implement and, as we will see, works surprising well.
The final exit decision is based on a simple threshold $\tau$ on the entropy. Specifically, $f^k(h^k)$ is 1 when $E^k < \tau$, and is otherwise 0.

\subsection{Training downstream tasks}

\begin{figure}
  \begin{center}
  \includegraphics[width=7cm]{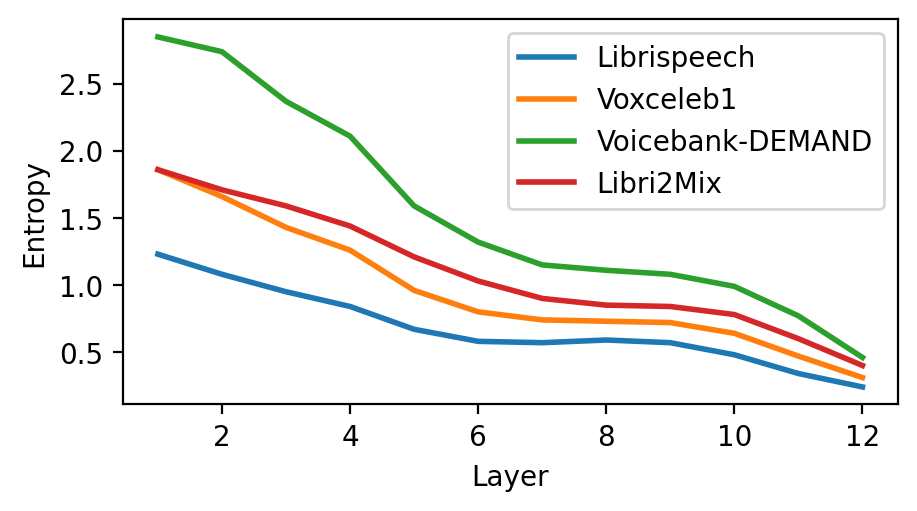}
  \end{center}
  \caption{The average entropy of each early exit branch on four datasets of MiniSUPERB \cite{wang2023minisuperb}.
   \label{fig:minisuperb-entropy}}
\end{figure}

To give a sense of how the entropy values vary across layers and across datasets,
we show the average entropy of each layer in HuBERT BASE on four datasets from MiniSUPERB \cite{wang2023minisuperb} in Figure \ref{fig:minisuperb-entropy}.
We observe that the entropy is smaller for deeper layers, and the entropy as a function of the layer is surpringly linear.\footnote{The linear relationship between the entropy and the layer is itself an interesting finding and warrants further study.}

The linear relationship between the entropy and the layers prompts us to design the following heuristic for choosing the threshold.
Given a dataset, we compute the average entropy of each layer and measure the maximum and minimum average entropy $E_{max}$ and $E_{min}$, respectively.
The threshold $\tau$ is decided by a linear scaling of the mean
\begin{align}
    \tau &= \frac{E_{max} + E_{min}}{2}\rho
\end{align}
where $\rho$ is a ratio hyperparameter in $[0,1]$.
The ratio $\rho$ is left to the user to decide; the smaller the $\rho$, the later the model exits and the more compute is required.

When training downstream models, given a threshold $\tau$, the model decides whether to exit at layer $k$ by evaluating $E^k < \tau$ for each data sample.
If early exit happens at layer $\hat{k}$, we apply Layer Normalization \cite{ba2016layer} on all hidden vectors below the $\hat{k}$-th layer,
\begin{align}
    \bar{h}^{1:\hat{k}} = LayerNorm(h^{1:\hat{k}}).
\end{align}
Following SUPERB \cite{shu2021superb}, the features to the downstream classifier is the weighted sum of the hidden vectors $\sum_{k=1}^{\hat{k}}w_k\bar{h^k}$.
The weights and the subsequent linear classifiers are trained for the downstream task.

\subsection{Exit strategies at inference time}

Though the hyperparameter $\rho$ needs to be fixed when training downstream models, it still can be changed during inference depending on how much compute the user is willing to spend, much like the spirit of anytime inference \cite{cai2020can}.
In fact, any additional constraints can be added at inference time, and below we explore three spans, constraining where the model can exit.
If during inference no layers within the span decides to exit early, we simply force the model to exit at the deepest layer within the defined span.

\vspace{-3mm}\paragraph*{mean} This constraint forces the model to exit at a layer that is the average number of layers used during training. Formally, if the average exit layer is $\mu$ during downstream training, we only allow the model to exit at layer $k$ if $\lfloor\mu\rfloor \leq k \leq \lceil\mu\rceil$. 
\vspace{-3mm}\paragraph*{threshold} This constraint only allows the model to exit at layers that have sufficiently been used as exit points during downstream training.  Formally, suppose the exit happens at the $k$-th layer $r_k$ of the time during downstream training. We only allow the model to exit at layer $k$, say, when $r_k > 15\%$. 
\vspace{-3mm}\paragraph*{min-max} This constraint imposes the minimum and maximum number of layers, and the two are determined based on the minimum and maximum used during downstream training. Formally, suppose the minimum exit layer is $k_\text{min}$ and the maximum is $k_\text{max}$ during downstream training. We only allow the model to exit at layer $k$ if $k_\text{min} \leq k \leq k_\text{max}$.  

%% file: experiments.tex
\section{Experiments}

We train the early exit branches on the full Librispeech 960 hours \cite{panayotov2015librispeech}. We fix all the other parameters and train the branches with a batch size of 32 and a learning rate of 5e-5. It takes 88000 steps to converge, about 10 hours on a single 32GB V100 GPU. We do not apply dropout and layer drop \cite{fan2019reducing} during training. All the other hyperparameters remain the same as HuBERT. We evaluate our models on MiniSUPERB \cite{wang2023minisuperb}, a lightweight benchmark that efficiently evaluate self-supervised speech models. There are four downstream tasks in MiniSUPERB, including automatic speech recognition (ASR), speaker identification (SID), speech enhancement (SE), and speech separation (SS). We use learning rate of 1e-4, 1e-1, 1e-4, and 1e-4 respectively for ASR, SID, SE, and SS.

%% file: results.tex
\subsection{Downstream performance and speed-up}

The downstream performance and the amount of forward time saved on MiniSUPERB \cite{wang2023minisuperb} are shown in Table \ref{tab:minisuperb}. 
In general, when using a smaller ratio $\rho$ (0.7), DAISY can achieve close or even better performance to HuBERT BASE on all four downstream tasks, while saving inference time. 
To highlight, on SID, it can improve accuracy by 1.21\% and save 23.36\% of forwarding time. On SE, it can improve PESQ by 0.02 and save 20.51\% of forwarding time.
When using a larger ratio $\rho$ (1.0), DAISY is able to save even more while still maintaining performance on SID, SE, and SS. 
To highlight, on SID, it can save 31.5\% of the forward time with less than 1\% accuracy degradation. On SE, it can save 29.56\% of forwarding time and improve PESQ by 0.02. On SS, it can save 24.75\% of forwarding time and significantly improve SI-SDRi by 0.19.
Compared to the other three tasks, DAISY finds it more challenging to save a substantial amount of forwarding time on ASR without sacrificing performance. This might be related to the distribution of layerwise information in self-supervised speech models, where phonetic information is predominantly stored in relatively deep layers~\cite{pasad2021layer}.

Comparing the three exit strategies at inference time, both mean and threshold methods can achieve a good trade-off between performance and forward time. We find that the min-max approach exhibits worse performance in a few cases.

\begin{figure}
  \centering
  \begin{center}
  \includegraphics[width=\linewidth]{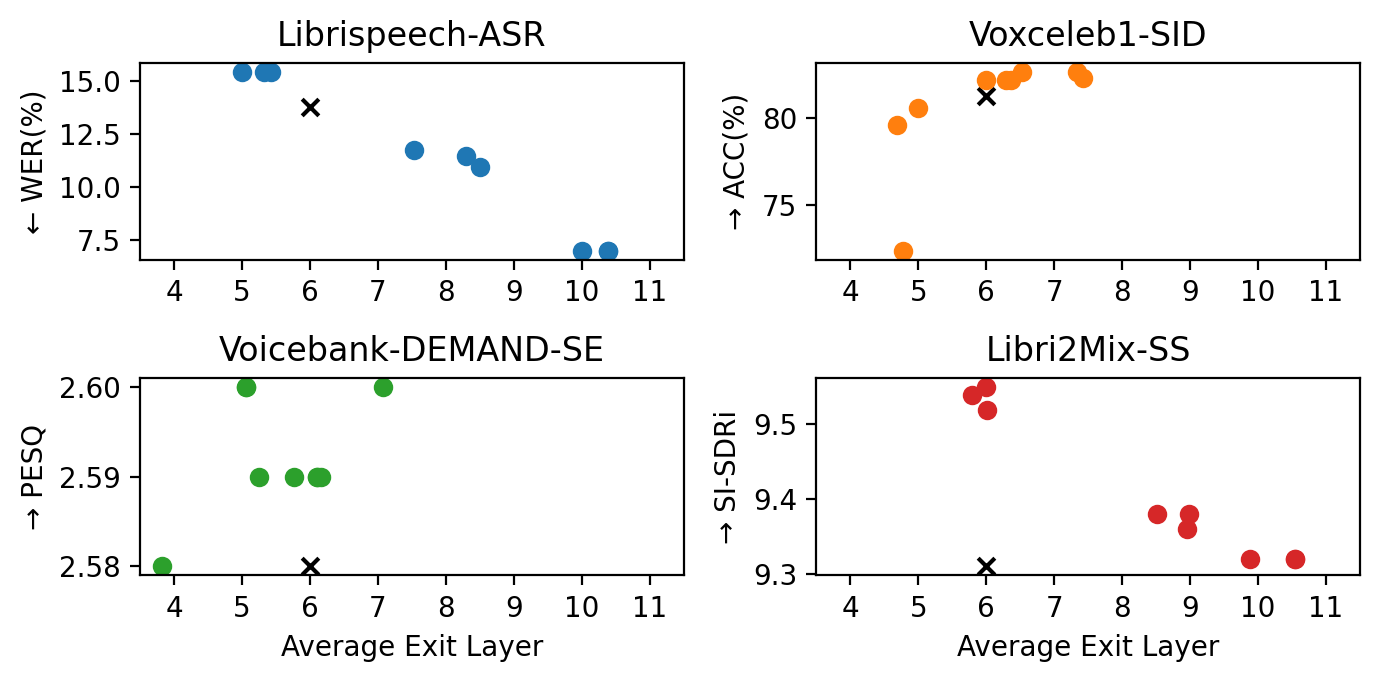}
  \end{center}
  \caption{The comparison of DAISY and the early exit baseline at a fixed layer. The colored dots represent DAISY, while the black crosses represent early exit at the 6th layer. For each downstream task, we present the results of three different $\rho$ values (0.7, 0.76, 1.0) combined with the three exit strategy at inference time, resulting in a total of 9 dots.  
   \label{fig:dynamic-static}}
\end{figure}

As Table \ref{tab:minisuperb} shows, users can decide the value of $\rho$ based on the amount of computational resources they have. When resources are limited, one can choose a larger $\rho$. 
In Figure \ref{fig:dynamic-static}, we further demonstrate the results of DAISY compared to exiting early at the 6th layer. We observe that DAISY is capable of achieving better performance than early exit at a fixed layer.
This shows the advantage of DAISY to dynamically determine the early exit layer.
In addition, using different $\rho$'s and different exit strategy provides a trade-off at inference time.

\begin{table*}
\caption{Downstream performance and the percentage of saving forward time of DAISY on four different downstream tasks of MiniSUPERB \cite{wang2023minisuperb}. ASR denotes automatic speech recognition, SID denotes speaker identification, SE denotes speech enhancement, and SS denotes speech separation. *The performance of HuBERT base 12 layer is copied from SUPERB benchmark \cite{shu2021superb}}
\centering
\vspace{1em}
\scalebox{0.78}{
\renewcommand{\arraystretch}{1.4}
\begin{tabular}{c|c|c|cc|cc|cc|cc}
 \hline
 \multicolumn{3}{c|}{} & \multicolumn{2}{c|}{ASR} & \multicolumn{2}{c|}{SID} & \multicolumn{2}{c|}{SE} & \multicolumn{2}{c}{SS} \\ \cline{4-11} 
 \multicolumn{3}{c|}{} & WER $\downarrow$ & Time Saved $\uparrow$ & ACC $\uparrow$ & Time Saved $\uparrow$ & PESQ / STOI $\uparrow$ & Time Saved $\uparrow$ & SI-SDRi $\uparrow$ & Time Saved $\uparrow$ \\ \hline
 \multirow{6}{*}{DAISY} & \multirow{3}{*}{$\rho=1.0$} & mean & 15.45\% & 21.63\% & 79.55\% & \textbf{31.51\%} & \textbf{2.60 / 94.0} & 27.23\% & 9.54 & 22.78\% \\ \cline{3-11} 
 & & threshold & 15.45\% & \textbf{31.54\%} & 80.53\% & 31.50\% & \textbf{2.60 / 94.0} & \textbf{29.56\%} & \textbf{9.55} & \textbf{24.75\%} \\ \cline{3-11} 
 &  & min-max & 15.43\% & 18.99\% & 72.36\% & 28.64\% & 2.58 / 93.9 & 29.22\% & 9.52 & 20.93\% \\ \cline{2-11} 
 & \multirow{3}{*}{$\rho=0.7$} & mean & 6.96\% & 4.98\% & \textbf{82.63\%} & 19.42\% & \textbf{2.60 / 94.0} & 20.51\% & 9.32 & 2.23\% \\ \cline{3-11} 
 &  & threshold & 6.98\% & 7.71\% & \textbf{82.63\%} & 23.36\% & 2.59 / 94.0 & 25.03\% & 9.32 & 2.50\% \\ \cline{3-11} 
 &  & min-max & 6.96\% & 4.74\% & 82.26\% & 15.19\% & 2.59 / 94.0 & 23.72\% & 9.32 & 1.25\% \\ \hline
 \multicolumn{3}{c|}{*HuBERT Base 12 layer \cite{hsu2021hubert}} & \textbf{6.42\%} & 0.00\% & 81.42\% & 0.00\% & 2.58 / 93.9 & 0.00\% & 9.36 & 0.00\% \\ \hline
\end{tabular}
}
\label{tab:minisuperb}
\end{table*}

%% file: analysis.tex
\begin{figure}[t]
  \begin{center}
  \includegraphics[width=7cm]{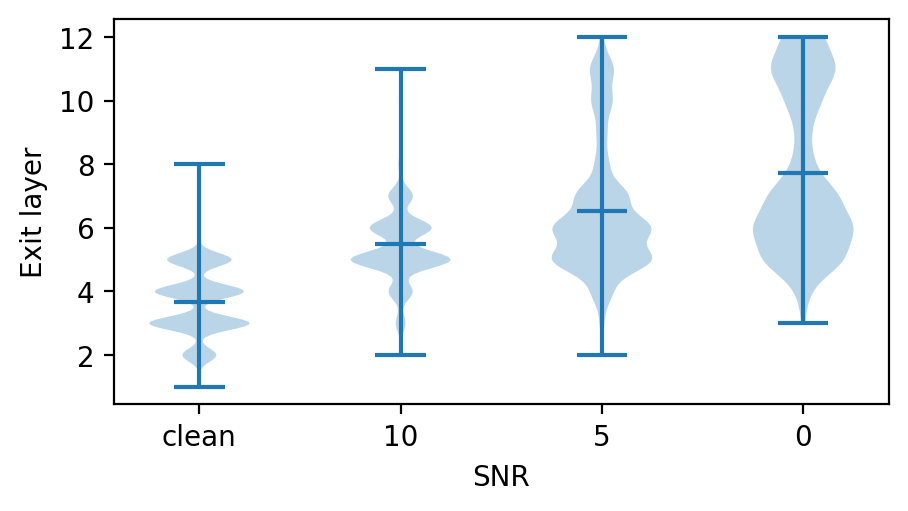} 
  \end{center}
  \caption{Violin plot of the early exit probability at each layer when applying different levels of MUSAN noise \cite{snyder2015musan} to the Librispeech test-clean set. The horizontal line represents the maximum, average, and minimum of the early exit layer, respectively. 
    \label{fig:snr-violin}}
  \vspace{-1em}
\end{figure}

\begin{figure}[t]
  \begin{center}
  \includegraphics[width=7cm]{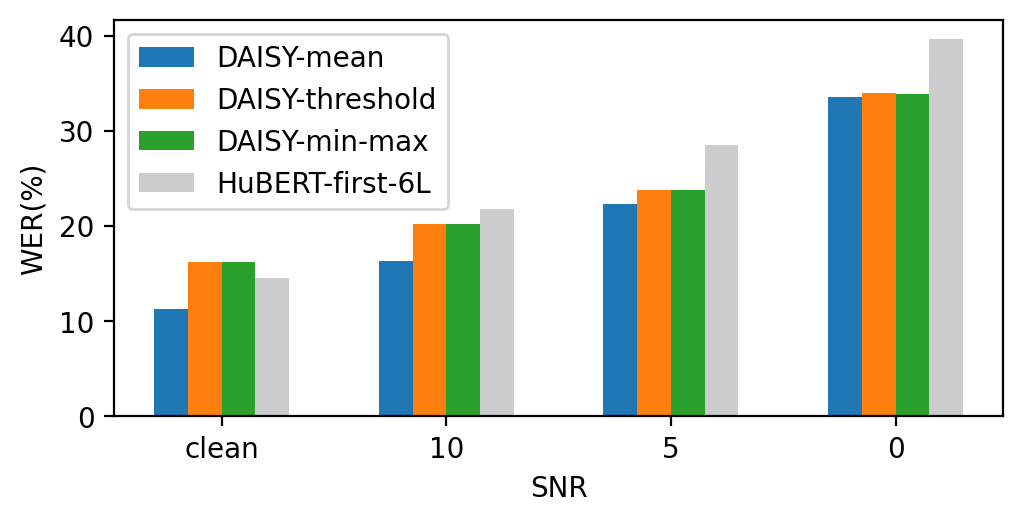}
  \end{center}
  \caption{The word error rate of DAISY on samples with different level of noises. HuBERT-first-6L represents the results of statically early exiting at 6th layer.
   \label{fig:snr-wer}}
   \vspace{-1.5em}
\end{figure}

\subsection{Noise adaptivity of DAISY}

From the results of the previous section, we can already see that DAISY is capable of dynamically deciding what samples require more layers to process and what samples do not.
However, it is unclear what properties of the data drive the early exit decision.
There might be many properties that determines when to exit.
In this section, we only focus on the noise level of the recordings, as the noise level can be controlled by adding noise to clean speech.
Specifically, we add different levels of noise to the Librispeech test clean dataset, including SNR values of 10, 5, and 0. We then measure the early exit results for each level of noise. The noise recordings were randomly sampled from the MUSAN dataset \cite{snyder2015musan}. The results are presented in Figure \ref{fig:snr-violin}. Overall, DAISY tends to use more layers as the SNR decreases.
Our findings is consistent with Ravuri et al.\ \cite{ravuri2023uncertainty}, where they found that the entropy of self-supervised speech models correlate well with their prediction (in their case, MOS scores). In addition to general background noise, we also experiment with adding music and speech, and the results are consistent with adding background noise.

\subsection{Applications of noise adaptativity}

The above analysis provides some evidence, supporting DAISY's ability to adapt to noise.
In particular, DAISY would have an advantage when the noise level has a large variance.
Unfortunately but also not surprisingly, most of the publicly available datasets have a relatively homogeneous noise level among data point.
To test the adaptivity of DAISY, we create a dataset with a range of noise levels.
We add MUSAN noise recordings to LibriSpeech 100 hours, dev clean, and test clean subsets, in which 40\% remain clean, 30\% of SNR 10, 20\% of SNR 5, and 10\% of SNR 0.
We then perform ASR on this dataset.
The result is shown in Figure \ref{fig:snr-wer}. Comparing to early exit at a fixed layer (the 6th), our approach performs better across all noise levels, and the gap becomes more pronounced when the level of noise increases. The average WER across all noise levels is 20.67, 23.44, 23.32, and 26.1 for DAISY-mean, DAISY-threshold, DAISY-min-max, and HuBERT-first-6L, respectively.
This result clearly shows that DAISY is able to identify the noisy samples, and, based on the noise level, allocate the appropriate amount of compute for this task.

Since most of the time we cannot precisely know how noisy a dataset is, and sometimes a dataset may contain data with varying degrees of noise, it becomes challenging to predetermine the appropriate layer for early exit apriori. In such cases, DAISY can dynamically adjust the amount of layer needed based on how noisy each sample is, thereby achieving better performance. 

%% file: conclusion.tex
\section{Conclusion}

In this work, we propose DAISY, a novel and effective approach for early exit based on a self-supervised loss. We use the entropy of the prediction to decide whether to exit early. Unlike previous early exit approaches, DAISY does not require training early exit branches for different downstream tasks and does not require any fine-tuning of the self-supervised model, making the training process much more efficient. On four downstream tasks of MiniSUPERB, DAISY achieves comparable or even better performance than HuBERT, while saving inference time. Lastly, we analyze the noise adaptivity property of DAISY, showing that noisy samples tend to require more layers to process. We then show that in a setting where the range of noise level is large, DAISY performs better comparing to early exit at a fixed layer, even though the amortized compute is the same.
The early exit approach provides a different avenue to model compression at achieving speed-up at inference time.  

%% file: adknowledgement.tex
\section{Acknowledgement}
We thank the National Center for High-performance
Computing (NCHC) of the National Applied Research Laboratories (NARLabs) in Taiwan for providing computational and storage resources.

%% file: main.bbl
\begin{thebibliography}{10}
\providecommand{\url}[1]{#1}
\csname url@samestyle\endcsname
\providecommand{\newblock}{\relax}
\providecommand{\bibinfo}[2]{#2}
\providecommand{\BIBentrySTDinterwordspacing}{\spaceskip=0pt\relax}
\providecommand{\BIBentryALTinterwordstretchfactor}{4}
\providecommand{\BIBentryALTinterwordspacing}{\spaceskip=\fontdimen2\font plus
\BIBentryALTinterwordstretchfactor\fontdimen3\font minus \fontdimen4\font\relax}
\providecommand{\BIBforeignlanguage}[2]{{%
\expandafter\ifx\csname l@#1\endcsname\relax
\typeout{** WARNING: IEEEtran.bst: No hyphenation pattern has been}%
\typeout{** loaded for the language `#1'. Using the pattern for}%
\typeout{** the default language instead.}%
\else
\language=\csname l@#1\endcsname
\fi
#2}}
\providecommand{\BIBdecl}{\relax}
\BIBdecl

\bibitem{chung2019unsupervised}
Y.-A. Chung, W.-N. Hsu, H.~Tang, and J.~Glass, ``An unsupervised autoregressive model for speech representation learning,'' in \emph{Interspeech}, 2019.

\bibitem{liu2020mockingjay}
A.~T. Liu, S.-w. Yang, P.-H. Chi, P.-c. Hsu, and H.-y. Lee, ``Mockingjay: {Unsupervised} speech representation learning with deep bidirectional transformer encoders,'' in \emph{ICASSP}.\hskip 1em plus 0.5em minus 0.4em\relax IEEE, 2020.

\bibitem{baevski2020wav2vec}
A.~Baevski, Y.~Zhou, A.~Mohamed, and M.~Auli, ``{wav2vec} 2.0: A framework for self-supervised learning of speech representations,'' in \emph{Advances in Neural Information Processing Systems}, 2020.

\bibitem{hsu2021hubert}
W.-N. Hsu, B.~Bolte, Y.-H.~H. Tsai, K.~Lakhotia, R.~Salakhutdinov, and A.~Mohamed, ``{HuBERT}: Self-supervised speech representation learning by masked prediction of hidden units,'' \emph{IEEE/ACM Transactions on Audio, Speech, and Language Processing}, 2021.

\bibitem{chen2022wavlm}
S.~Chen, C.~Wang, Z.~Chen, Y.~Wu, S.~Liu, Z.~Chen, J.~Li, N.~Kanda, T.~Yoshioka, X.~Xiao \emph{et~al.}, ``{WavLM}: Large-scale self-supervised pre-training for full stack speech processing,'' \emph{IEEE Journal of Selected Topics in Signal Processing}, 2022.

\bibitem{chung2021w2v}
Y.-A. Chung, Y.~Zhang, W.~Han, C.-C. Chiu, J.~Qin, R.~Pang, and Y.~Wu, ``{w2v-BERT}: Combining contrastive learning and masked language modeling for self-supervised speech pre-training,'' in \emph{ASRU}, 2021.

\bibitem{shu2021superb}
S.-w. Yang, P.-H. Chi, Y.-S. Chuang, C.-I.~J. Lai, K.~Lakhotia, Y.~Y. Lin, A.~T. Liu, J.~Shi, X.~Chang, G.-T. Lin \emph{et~al.}, ``{SUPERB}: Speech processing universal performance benchmark,'' in \emph{Interspeech}, 2021.

\bibitem{tsai2022superb}
H.-S. Tsai, H.-J. Chang, W.-C. Huang, Z.~Huang, K.~Lakhotia, S.-w. Yang, S.~Dong, A.~T. Liu, C.-I.~J. Lai, J.~Shi \emph{et~al.}, ``{SUPERB-SG: Enhanced speech processing universal performance benchmark for semantic and generative capabilities},'' \emph{ACL}, 2022.

\bibitem{2022superbslt}
anonymous, ``{SUPERB @ SLT 2022}: Challenge on generalization and efficiency of self-supervised speech representation learning,'' in \emph{SLT}, 2022.

\bibitem{zhang2022bigssl}
Y.~Zhang, D.~S. Park, W.~Han, J.~Qin, A.~Gulati, J.~Shor, A.~Jansen, Y.~Xu, Y.~Huang, S.~Wang \emph{et~al.}, ``{BigSSL}: Exploring the frontier of large-scale semi-supervised learning for automatic speech recognition,'' \emph{IEEE Journal of Selected Topics in Signal Processing}, 2022.

\bibitem{chang2022distilhubert}
H.-J. Chang, S.-w. Yang, and H.-y. Lee, ``{DistilHuBERT}: Speech representation learning by layer-wise distillation of hidden-unit {BERT},'' in \emph{ICASSP}, 2022.

\bibitem{wang2022lighthubert}
R.~Wang, Q.~Bai, J.~Ao, L.~Zhou, Z.~Xiong, Z.~Wei, Y.~Zhang, T.~Ko, and H.~Li, ``{LightHuBERT}: Lightweight and configurable speech representation learning with once-for-all hidden-unit {BERT},'' in \emph{Interspeech}, 2022.

\bibitem{lee2022fithubert}
Y.~Lee, K.~Jang, J.~Goo, Y.~Jung, and H.~Kim, ``{FitHuBERT}: Going thinner and deeper for knowledge distillation of speech self-supervised learning,'' in \emph{Interspeech}, 2022.

\bibitem{lai2021parp}
C.-I.~J. Lai, Y.~Zhang, A.~H. Liu, S.~Chang, Y.-L. Liao, Y.-S. Chuang, K.~Qian, S.~Khurana, D.~Cox, and J.~Glass, ``{PARP}: Prune, adjust and re-prune for self-supervised speech recognition,'' in \emph{Advances in Neural Information Processing Systems}, 2021.

\bibitem{peng2023structured}
Y.~Peng, K.~Kim, F.~Wu, P.~Sridhar, and S.~Watanabe, ``Structured pruning of self-supervised pre-trained models for speech recognition and understanding,'' in \emph{ICASSP}, 2023.

\bibitem{wang2023task}
H.~Wang, S.~Wang, W.-Q. Zhang, H.~Suo, and Y.~Wan, ``Task-agnostic structured pruning of speech representation models,'' in \emph{Interspeech}, 2023.

\bibitem{peng2021shrinking}
Z.~Peng, A.~Budhkar, I.~Tuil, J.~Levy, P.~Sobhani, R.~Cohen, and J.~Nassour, ``Shrinking bigfoot: Reducing {wav2vec} 2.0 footprint,'' in \emph{Workshop on Simple and Efficient Natural Language Processing}, 2021.

\bibitem{teerapittayanon2016branchynet}
S.~Teerapittayanon, B.~McDanel, and H.-T. Kung, ``{BranchyNet: Fast inference via early exiting from deep neural networks},'' in \emph{ICPR}, 2016.

\bibitem{xin2020deebert}
J.~Xin, R.~Tang, J.~Lee, Y.~Yu, and J.~Lin, ``{DeeBERT: Dynamic Early Exiting for Accelerating BERT Inference},'' \emph{ACL}, 2020.

\bibitem{zhou2020bert}
W.~Zhou, C.~Xu, T.~Ge, J.~McAuley, K.~Xu, and F.~Wei, ``{BERT} loses patience: Fast and robust inference with early exit,'' in \emph{Advances in Neural Information Processing Systems}, 2020.

\bibitem{xin2021berxit}
J.~Xin, R.~Tang, Y.~Yu, and J.~Lin, ``{BERxiT: Early Exiting for BERT with Better Fine-Tuning and Extension to Regression},'' in \emph{Proceedings of the 16th conference of the European chapter of the association for computational linguistics: Main Volume}, 2021.

\bibitem{li2021accelerating}
X.~Li, Y.~Shao, T.~Sun, H.~Yan, X.~Qiu, and X.~Huang, ``Accelerating bert inference for sequence labeling via early-exit,'' \emph{arXiv preprint arXiv:2105.13878}, 2021.

\bibitem{schuster2022confident}
T.~Schuster, A.~Fisch, J.~Gupta, M.~Dehghani, D.~Bahri, V.~Tran, Y.~Tay, and D.~Metzler, ``{Confident Adaptive Language Modeling},'' \emph{NeurIPS}, 2022.

\bibitem{hu2023smartbert}
B.~Hu, Y.~Zhu, J.~Li, and S.~Tang, ``{SmartBERT: A Promotion of Dynamic Early Exiting Mechanism for Accelerating BERT Inference},'' \emph{IJCAI}, 2023.

\bibitem{yoon2022hubert}
J.~W. Yoon, B.~J. Woo, and N.~S. Kim, ``{HuBERT-EE: Early Exiting HuBERT for Efficient Speech Recognition},'' \emph{arXiv preprint arXiv:2204.06328}, 2022.

\bibitem{lin2022compressing}
T.-Q. Lin, T.-H. Yang, C.-Y. Chang, K.-M. Chen, T.-h. Feng, H.-y. Lee, and H.~Tang, ``{Compressing Transformer-based self-supervised models for speech processing},'' \emph{arXiv preprint arXiv:2211.09949}, 2022.

\bibitem{yang2023fast}
G.~Yang, Z.~Ma, Z.~Zheng, Y.~Song, Z.~Niu, and X.~Chen, ``{Fast-HuBERT: An Efficient Training Framework for Self-Supervised Speech Representation Learning},'' \emph{ASRU}, 2023.

\bibitem{ovadia2019can}
Y.~Ovadia, E.~Fertig, J.~Ren, Z.~Nado, D.~Sculley, S.~Nowozin, J.~V. Dillon, B.~Lakshminarayanan, and J.~Snoek, ``Can you trust your model's uncertainty? evaluating predictive uncertainty under dataset shift,'' in \emph{Advances in Neural Information Processing Systems}, 2019.

\bibitem{wang2023minisuperb}
Y.-H. Wang, H.-Y. Chen, K.-W. Chang, W.~Hsu, and H.-y. Lee, ``Minisuperb: Lightweight benchmark for self-supervised speech models,'' \emph{ASRU}, 2023.

\bibitem{ba2016layer}
J.~L. Ba, J.~R. Kiros, and G.~E. Hinton, ``Layer normalization,'' \emph{arXiv preprint arXiv:1607.06450}, 2016.

\bibitem{cai2020can}
H.~Cai, C.~Gan, T.~Wang, Z.~Zhang, and S.~Han, ``Once-for-all: Train one network and specialize it for efficient deployment,'' in \emph{ICLR}, 2020.

\bibitem{panayotov2015librispeech}
V.~Panayotov, G.~Chen, D.~Povey, and S.~Khudanpur, ``Librispeech: an asr corpus based on public domain audio books,'' in \emph{ICASSP}, 2015.

\bibitem{fan2019reducing}
A.~Fan, E.~Grave, and A.~Joulin, ``Reducing transformer depth on demand with structured dropout,'' \emph{arXiv preprint arXiv:1909.11556}, 2019.

\bibitem{pasad2021layer}
A.~Pasad, J.-C. Chou, and K.~Livescu, ``Layer-wise analysis of a self-supervised speech representation model,'' in \emph{ASRU}, 2021.

\bibitem{snyder2015musan}
D.~Snyder, G.~Chen, and D.~Povey, ``Musan: A music, speech, and noise corpus,'' \emph{arXiv preprint arXiv:1510.08484}, 2015.

\bibitem{ravuri2023uncertainty}
A.~Ravuri, E.~Cooper, and J.~Yamagishi, ``Uncertainty as a predictor: Leveraging self-supervised learning for zero-shot mos prediction,'' \emph{arXiv preprint arXiv:2312.15616}, 2023.

\end{thebibliography}
